\documentclass[letterpaper, 10 pt, conference]{ieeeconf}  % Comment this line out if you need a4paper
\usepackage{amsmath} % align(ed)
\usepackage{amsfonts} % \mathbb, ...
\usepackage{amssymb} % for \triangleq
\usepackage{graphicx}
\usepackage{multicol} % For showing a multi-column figure
\usepackage{algorithm} % For the pseudo-algorithm
\usepackage{algpseudocode}
\usepackage[dvipsnames]{xcolor} % For setting text color
\usepackage{multirow}
\usepackage[hang,flushmargin]{footmisc} % To remove indentation in footnotes
\usepackage{booktabs} % \toprule \bottomrule for tables
\usepackage{mathtools}

\usepackage{caption} % For using subfigure
\usepackage{subcaption} % For using subfigure

\newcommand*\subtxt[1]{_{\mathrm{#1}}}
\DeclareRobustCommand\_{\ifmmode\expandafter\subtxt\else\textunderscore\fi}

\newcommand{\ts}{\textsuperscript}

\newcommand{\pderiv}[2]{\frac{\partial #1}{\partial #2}}
\newcommand{\pderivv}[2]{\frac{\partial^{2} #1}{\partial #2^{2}}}

\newcommand{\pderivil}[2]{{\partial #1}/{\partial #2}}

\newtheorem{assumption}{Assumption}
\newtheorem{remark}{Remark}

\IEEEoverridecommandlockouts                              % This command is only needed if 
\title{\LARGE \bf
Analysis of extremum seeking control for wind turbine torque controller optimization by aerodynamic and generator power objectives}

\author{S.P. Mulders$^{1}$, A.J. Gallo$^{1}$, M.A. Rotea$^{2}$ % <-this % stops a space  ,M.A. Rotea$^{2}$
\thanks{$^{1}$Delft University of Technology, Delft Center for Systems and Control, Mekelweg 2, 2628 CD Delft, The Netherlands. {\tt\small \{S.P.Mulders,\, A.J.Gallo\}@tudelft.nl}}
\thanks{$^{2}$University of Texas at Dallas, Center for Wind Energy and Department of Mechanical Engineering, Richardson, TX 75080, USA. {\tt\small {rotea@utdallas.edu}}}
\thanks{~~This research work has been funded in part by Analytics for asset Integrity Management of Windfarms (AIMWind), under grant no. 312486, from Research Council of Norway (RCN). AIMWind is a collaborative research from University of Agder, Norwegian Research Center (NORCE), and TU Delft, with Origo Solutions as advisory partner.}%
}

\begin{document}

\maketitle
\thispagestyle{empty}
\pagestyle{empty}

\bstctlcite{IEEEexample:BSTcontrol}

%%%%%%%%%%%%%%%%%%%%%%%%%%%%%%%%%%%%%%%%%%%%%%%%%%%%%%%%%%%%%%%%%%%%%%%%%%%%%%%%
\begin{abstract}
Wind turbines degrade over time, resulting in varying structural, aeroelastic, and aerodynamic properties. In contrast, the turbine controller calibrations generally remain constant, leading to suboptimal performance and potential stability issues. The calibration of wind turbine controller parameters is therefore of high interest. To this end, several adaptive control schemes based on extremum seeking control (ESC) have been proposed in the literature. These schemes have been successfully employed to maximize turbine power capture by optimization of the $K\omega^2$-type torque controller. In practice, ESC is performed using electrical generator power, which is easily obtained. This paper analyses the feasibility of torque gain optimization using aerodynamic and generator powers. It is shown that, unlike aerodynamic power, using the generator power objective limits the dither frequency to lower values, reducing the convergence rate unless the phase of the demodulation ESC signal is properly adjusted. A frequency-domain analysis of both systems shows distinct phase behavior, impacting ESC performance. A solution is proposed by constructing an objective measure based on an estimate of the aerodynamic power.
\end{abstract}
%%%%%%%%%%%%%%%%%%%%%%%%%%%%%%%%%%%%%%%%%%%%%%%%%%%%%%%%%%%%%%%%%%%%%%%%%%%%%%%%
\section{INTRODUCTION}
\noindent The increasing need to accelerate the global transition toward renewable energy sources motivates the development of ever-larger wind turbines with increased power capacities. The upscaling of wind turbines leads to a taller support structure and increased rotor sizes with higher flexibility, bringing challenges for structural and controller design.

Conventional and advanced turbine controller strategies often rely on model accuracy due to the limited measurements available to a turbine controller. As shown in~\cite{Brandetti2022}, model parameter inaccuracies result in turbine operation away from the intended operating point, potentially causing suboptimal operational behavior and possibly leading to stability issues. Such discrepancies can become more significant over time due to aerodynamic degradation of the rotor by, e.g., wear and tear, bug build-up, and icing~\cite{Johnson2006}. As often the relation between turbine degradation and physical turbine properties is a priori unknown, learning schemes capable of online tuning of control systems, through the calibration of internal (physical) model information, are currently of high interest~\cite{Mulders2023_ACC,Mulders2023_IFAC}.

For present-day multi-megawatt turbines, the conventional and (relatively) straightforward $K\omega^2$  ({"K-omega-squared"}) torque control strategy still shows good performance regarding power extraction for modern large-scale wind turbines~\cite{Brandetti2023_MOControlCalibration}. Although the performance of this controller type highly depends on the quality of the model information it is based on, the single torque gain structure allows for convenient direct optimization. In the past decade, numerous works have been published on this aspect, proposing extremum seeking control (ESC) as a viable candidate for real-world and real-time controller optimization. 

ESC is an adaptive control algorithm that optimizes steady-state input-output mappings of (dynamic) systems that possess (local) optima~\cite{Ariyur2003_ESCBook, Rotea2001_ACC, Tan2010_ESC1922to2010}. The ESC algorithm is model-free, does not require prior system knowledge, and is based on the notion of time-scale separation. The algorithm extracts information on the gradient of some measurable objective cost to the optimized variable by periodically exciting the system through a dither signal and subsequently manipulating output signals. The algorithm can be applied to optimize time-invariant or slowly time-varying systems.

ESC for wind turbine controller optimization was initially employed to optimize the constant pitch angle for turbine energy capture maximization~\cite{Komatsu2001_ESCFinePitch}. A similar implementation was later evaluated on a laboratory-scale two-bladed micro wind turbine~\cite{Ishii2003_ESCMicroWindTubine}. The scheme was extended in a multivariable setting by optimizing the combined torque and pitch to maximize energy capture~\cite{Creaby2009_MultiVarESC}, the effectiveness of which was later evaluated on a simulation model of the National Renewable Energy Laboratory (NREL) Controls Advanced Research Turbine (CART3) turbine~\cite{Xiao2016_CART3Sim} and in field-test on the actual CART3 turbine~\cite{Xiao2018_CART3Field}. Further works improved the ESC algorithm's convergence, making it uncorrelated to the mean wind speed by exploiting the logarithm of the power signal as objective~\cite{Rotea2017_LogofPowers}. This latter improvement was validated with full large eddy simulations in~\cite{Ciri2019_LESLPESC}. Later in \cite{Kumar2022LPPIESC}, the convergence of this algorithm was accelerated at the expense of introducing additional tuning parameters.

Industrial turbines seldom have the means to measure the aerodynamic torque or power directly; nevertheless, wind turbine generator power is accurately measured. Assuming ideal drivetrain efficiency, aerodynamic and generator power are equal in steady state. Thus, the generator power measurement is a natural optimization objective candidate. However, from exploratory numerical simulations, it was observed by the authors of this paper that the adoption of conventional ESC based on generator power-based optimization posed challenges. Its use might still be feasible with lower dither frequencies but with increased complexity in calibration and potential increased convergence times. The underlying cause of this phenomenon has never been described in the open literature and is the motivation for the analysis presented in this paper.
%as the rotor inertia increases.

This paper analyses the dynamic properties of the system that is subject to dither-demodulation ESC with torque gain as optimization input and measured aerodynamic and generator power signals as objective output. It has been found that the system dynamics are different. These dynamic differences impact the application of a dither-demodulation ESC with measured generator power as an optimization objective. This paper thereby presents the following contributions:
\begin{enumerate}
    \item Describing the implications when implementing generator power-based ESC by simulations.
    \item Providing a dynamical analysis for both aerodynamic power and generator power as ESC objectives.
    \item Proposing a solution that improves ESC convergence by formulating a new estimated aerodynamic power objective based on augmented measured generator power with rotor acceleration dynamics.
\end{enumerate}
In this paper, the issue is presented and an intuitive explanation of its cause is provided. Specifically, a dynamic analysis based on frequency-domain analysis via linearizations is proposed.
% of the dynamics, demonstrating the root of the problem.

The paper is organized as follows. Sections~\ref{sec:T}~and~\ref{sec:ESC} establish the wind turbine model and the employed ESC scheme. Section~\ref{sec:PI} describes torque gain optimization based on both power objectives.  Section~\ref{sec:DA} presents a dynamical analysis of the systems considered for operating points around the optimal torque gain value. Section~\ref{sec:S} proposes a solution for faster ESC optimization by estimating the aerodynamic power with an augmented generator power objective. Finally, conclusions are drawn in Section~\ref{sec:C}.

\section*{ASSUMPTIONS AND PREREQUISITES}
\noindent The following assumptions are used in this paper:
\begin{assumption}\label{ass:DrivetrainPerfect}
    The drive-train is considered direct-drive, perfectly stiff, and thereby static (nondynamic). Its efficiency and the efficiency of the electrical generator are ideal.
\end{assumption}
\begin{assumption}\label{ass:RotorSpeed}
    Rotor speed and generator power can be measured without being corrupted by noise.
\end{assumption}
\begin{assumption}\label{ass:WindSpeed}
    Constant, below-rated wind speed is considered for convergence analysis.
    % A single, constant wind speed is studied for analysis; the conclusions drawn are equally valid at other turbine operating points in partial load.
\end{assumption}
\noindent Note that, in addition to Assumption~\ref{ass:WindSpeed}, all the plots shown throughout this paper are considered with a single constant wind speed of $8$~m/s. 
%This is done without loss of generality, and the conclusions drawn are equally valid at other turbine operating conditions in partial load.
Furthermore, throughout this paper, a direct-drive turbine (gearbox ratio of one) with the aerodynamic properties of the {NREL~5\nobreakdash-MW} reference turbine is taken. The classical dither-demodulation ESC scheme for real-time optimization is considered.

Values corresponding to a specific operating point are denoted by $\bar{\left(\cdot\right)}$. Signal time-derivatives are indicated by $\dot{\left(\cdot\right)}$, whereas estimated quantities are represented by $\hat{\left(\cdot\right)}$, and the notation ${\left(\cdot\right)^{*}}$ indicates an optimal value.

\section{WIND TURBINE MODEL}\label{sec:T}
\noindent Most modern wind turbines employ a partial-load torque control strategy, maximizing wind power capture by operating around peak aerodynamic rotor efficiency. This section defines aerodynamic relations to establish the considered torque controller strategy and a dynamic wind turbine model. %For lower wind speed, wind turbines operate in the so-called partial-load conditions.

\subsection{Aerodynamic relations}\label{sec:T_SS}
\noindent The available wind power in the rotor-swept surface area is given by
\begin{align}
    P\_{w} &= \frac{1}{2}\rho A V^{3},
    \label{eq:T_WindPower}
\end{align}
with $\rho\in\mathbb{R}^\mathrm{+}$ being the fluid (air) density, $A=\pi R^2$ the rotor swept area with $R\in\mathbb{R}^\mathrm{+}$ the rotor radius and $V\in\mathbb{R}^\mathrm{+}$ the rotor-effective wind speed. 

The fraction of power that is extracted in steady-state by the wind turbine rotor from the available wind power is described by an operating condition-dependent power coefficient ${C\_{p}(\lambda,\beta):\mathbb{R}\times\mathbb{R}\rightarrow\mathbb{R}}$. This nonlinear and rotor-design dependent mapping represents the rotor aerodynamic efficiency as a function of the tip-speed ratio $\lambda$ and pitch angle $\beta$. The tip-speed ratio is further defined as
\begin{align}
    \lambda = \frac{\omega\_{r}R}{V},
    \label{eq:T_TSR}
\end{align}
indicating the ratio between the blade tip speed and incoming effective wind speed, where $\omega\_{r}$ is the rotor angular velocity. The aerodynamic power is, in turn, defined as
\begin{align}
    P\_{r} &= \tau\_{r}\omega\_{r} = C\_{p}(\lambda,\beta)P\_{w}.
    \label{eq:T_RotorPower}
\end{align}
An explicit relation for the rotor torque is given as
\begin{align}
    \tau\_{r} &= \frac{1}{2}\rho\_{a}ARU^2 C\_{\tau}(\lambda, \beta), 
\end{align}
where the torque coefficient is related to the power coefficient as
\begin{align*}
    C\_{\tau}(\lambda,\beta) &= {C\_{p}}(\lambda,\beta)/{\lambda}.
\end{align*}
The next section employs the relations in this section to establish the dynamic wind turbine model.

\subsection{Dynamic wind turbine model}\label{sec:T_WT}
\noindent The wind turbine aerodynamics are modeled through a nonlinear first-order system
\begin{align}
I\dot{\omega}\_{r} &= \tau\_{r} - \tau\_{g},
\label{eq:T_FirstOrderAeroDiffEq}
\end{align}
in which $I\in\mathbb{R}^{+}$ is the combined rotor and drivetrain inertia, and $\tau\_{g}\in\mathbb{R}$ is the generator torque. In this work, the actual generator torque is assumed to equal its set point (Assumption~\ref{ass:DrivetrainPerfect}). The static $K\omega^2$ feedback controller generates the generator torque set point
\begin{align}
    \tau\_{g} &= K\omega\_{r}^2,
    \label{eq:T_Kw3Power}
\end{align}
which tracks a predefined operating point in steady state, where the torque gain $K$ is computed as
\begin{align}
    K(C\_p(\lambda,\beta)) = \frac{\pi\rho R^5 C\_{p}(\lambda,\beta)}{2\lambda^{3}}.
    \label{eq:T_TorqueGainK}
\end{align}
To maximize turbine power production, the tip-speed ratio and pitch angle are chosen as $\lambda^{*}$ and $\beta^{*}$ maximizing the power coefficient $C\_{p}^{*} = C\_{p}(\lambda^{*},\beta^{*})$ resulting in the optimal torque gain $K^{*}=K(C\_{p}^{*})$. Finally, the measured generator power is defined as
\begin{align}
    P\_{g} &= \tau\_{g}\omega\_{r} = K\omega\_{r}^3.
\end{align}
Now that the aerodynamic relations, dynamic turbine model, and torque control strategy have been formulated, the next section summarizes the ESC algorithm and implementation.

\section{EXTREMUM SEEKING CONTROL}\label{sec:ESC}
\noindent This section summarizes the dither-demodulation extremum seeking control algorithm and describes its use for turbine torque control optimization for power capture maximization. An explicit definition of the ESC dynamics is left out of this work for reasons of brevity; the reader is referred to~\cite{Rotea2001_ACC, Ariyur2003_ESCBook,Tan2010_ESC1922to2010} for a more elaborate definition and explanation of the algorithm.

\subsection{Algorithm description}\label{sec:ESC_Alg}
\begin{figure}[b!]
    \centering
    \includegraphics[scale=1.0, trim=0cm 0cm 0cm 0cm, clip]{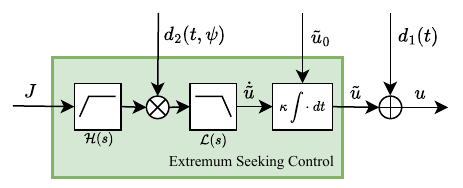}
    \caption{Dither-demodulation extremum seeking control scheme. The reader is referred to the main text for interpretation of this schematic.}
    \label{fig:ESCAlgBlock}
\end{figure}
\noindent A schematic representation of the considered dither-demodulation ESC scheme is presented in Figure~\ref{fig:ESCAlgBlock}. The goal of the algorithm is to find a (local) optimum of the cost $J$ by estimating its gradient to the control parameter $u$, with its time-derivative defined as
\begin{align}
    \dot{\tilde{u}} = \kappa \pderiv{J}{u},
\end{align}
where $\dot{\tilde{u}}$ is the time-derivative of the control parameter under optimization, and $\kappa$ is the convergence gain of the algorithm. Next, a brief nonanalytic explanation is given on how the scheme retrieves the gradient $\pderivil{J}{u}$.

As shown in Figure~\ref{fig:ESCAlgBlock}, the measured performance objective $J$ is high-pass filtered by $\mathcal{H}(s, \omega_{\mathcal{H}})$ to remove the DC-offset. It is subsequently demodulated by multiplication with the demodulation signal $d_2(t,\psi) = \sin{(\omega\_{d}t+\psi)}$ of unity amplitude, and with dither frequency $\omega\_{d}$. The dither frequency may be phase-compensated by $\psi$ to account for the system and high-pass filter's phase loss, improving convergence characteristics. This operation results in gradient information as a steady-state contribution and at twice the dither frequency. The low-pass filter $\mathcal{L}(s, \omega_{\mathcal{L}})$ attenuates noise and the higher harmonic dither frequency, resulting in $\dot{\tilde{u}}$ of which the steady-state component represents the desired gradient~$\pderivil{J}{u}$. The signal $\dot{\tilde{u}}$ is integrated with respect to time and scaled by $\kappa$ for convergence. The respective high- and low-pass filters' cut-in and cut-off frequencies are related to the dither frequency by
\begin{align}
    \omega_{\mathcal{H}} = \omega\_{d}/5,\quad \omega_{\mathcal{L}} = 2\omega\_{d}.\label{eq:ESC_Alg_FilterW}
\end{align}
The scaled dither signal $d_1(t)=A\_{d}\sin{(\omega\_{d}t)}$ is added to ${\tilde{u}}$ and forms the combined steady-state and periodic input $u$ to the system under optimization. The dither amplitude $A\_{d}$ is calibrated as a trade-off between excitation level (signal-to-noise ratio, SNR) and system limitations.

This real-time process proceeds until the gradient diminishes and the control input converges to a neighborhood of the optimum $\tilde{u}^*$.
Specifically, the ESC algorithm converges to a neighborhood of the optimal control parameter $u^{*}$, an optimizer for the cost function~\cite{Ariyur2003_ESCBook}, whenever the following three conditions are met\begin{subequations}
    \begin{align}
        &\pderiv{J}{u}(u^{*}) = 0~\text{(optimum exists)},\label{eq:ESC_Cond1}\\
        &\pderivv{J}{u}(u^{*}) < 0~\text{(maximum)},\label{eq:ESC_Cond2}\\
        &\pderiv{J}{u}(u^{*}+\zeta)\zeta < 0~\forall~\zeta\in Z,\label{eq:ESC_Cond3}
    \end{align}
\end{subequations}
where~\eqref{eq:ESC_Cond3} indicates that $u^*$ is a local maximizer in some region $Z$.

\subsection{ESC for torque control optimization}\label{sec:ESC_IMPL}
\begin{figure}[t!]
    \centering
    \includegraphics[scale=1.0, trim=0mm 0cm 0cm 0cm]{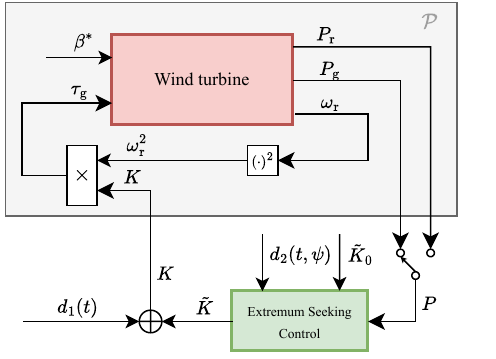}
    \caption{ESC for torque gain optimization for a $K\omega^2$ type of turbine torque controller. Two objectives, aerodynamic power and generator power, are considered for analysis.}
    \label{fig:ESCBlockSchemeK}
\end{figure}
\noindent This section describes the ESC algorithm's implementation for torque controller optimization and is schematically presented in Figure~\ref{fig:ESCBlockSchemeK}. The control variable that is considered for optimization is the torque gain ${u=K}$. The torque gain is integrated into the ESC scheme, where the integrator is initialized with the initial condition $\tilde{u}_0 = \tilde{K}_0$. The torque gain is periodically excited by an amplitude-scaled dither signal, such that ${K = \tilde{K}+d_1(t)}$. The performance objective ${J = P = \left\{P\_{r},\,P\_{g}\right\}}$ to the ESC algorithm is either the aerodynamic power $P\_{r}$ or the generator power $P\_{g}$. It has been shown that, by selecting the torque gain $K$, the WT steady state power output can be maximized \cite{Johnson2006}, indicating that WT power is, in steady state, concave in the control variable $K$. The single-input multiple-output (SIMO) system -- included in the gray box in Figure~\ref{fig:ESCBlockSchemeK} -- is referred to as $\mathcal{P}$ in the sequel of this paper. The next section provides a problem illustration for ESC torque controller optimization based on both objectives.

\section{PROBLEM ILLUSTRATION}\label{sec:PI}
\begin{figure}[b!]
    \centering
    \includegraphics[scale=1.0,trim={0cm 0cm 0cm 0cm}, clip]{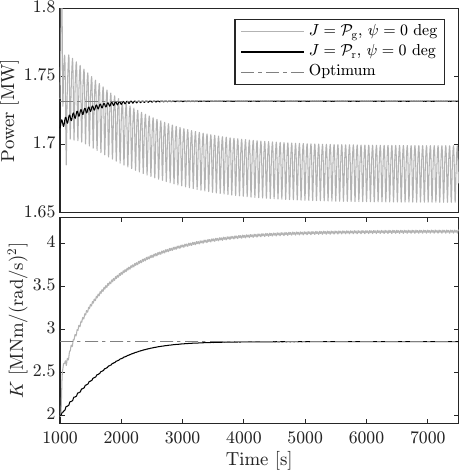}
    \caption{Torque gain optimization on aerodynamic power ($J=\mathcal{P}_\mathrm{r}$) and generator power ($J=\mathcal{P}_\mathrm{g}$) objectives, without demodulation phase compensation ($\psi=0$). ESC optimization based on aerodynamic power shows convergence to the actual optimum, whereas generator power-based optimization shows convergence to a nonoptimal value.}
    \label{fig:E_ProblemIllustration1}
    % \caption{Problem illustration comparing ESC with aerodynamic power and generator power as optimization objective. For both cases, the excitation amplitude and frequency are chosen equal, and the convergence gain is adapted to match the convergence speed between both cases. It is immediately apparent that while using aerodynamic power as the performance objective results in convergence; using measured generator power does not converge and surpasses the optimal value.}
\end{figure}
\noindent As discussed earlier, aerodynamic torque and power are rarely measured directly in commercial turbines and, under Assumption~\ref{ass:DrivetrainPerfect}, aerodynamic and generator powers are equal in steady state. Therefore, generator power measurements are compelling for the practical implementation of ESC. This section provides illustrative simulations of different outcomes that may occur when using the dither-demodulation ESC algorithm presented in the previous section, based either on measured aerodynamic or generator power.

Specifically, it is shown that, when using the generator power as the ESC objective function, the phase compensation $\psi$ in the demodulation signal $d_2$ must be tuned precisely for higher dither frequencies to achieve comparable convergence properties to the case in which aerodynamic power is used. In contrast, ESC with aerodynamic power as the objective demonstrates robustness to the choice of $\psi$. Additionally, it is shown that by reducing the dither frequency, ESC with generator power as the objective shows increasing robustness to phase compensation calibrations.

Three simulation scenarios are given for illustration purposes. In the first, the two ESC schemes are shown without tuning the phase offset while using the same dither frequency. In the second, phase compensation is introduced, showing the aerodynamic power's robustness and the sensitivity of the generator power objective to the specific choice of $\psi$. The third scenario shows the effect of lowering the dither signal. The ESC algorithm is implemented on both objectives and subject to a wind speed of $8$~m/s, and a dither amplitude $A\_{d} = 1\cdot10^{5}$~Nm/(rad/s)\ts{2} is used for all cases.

The results of the first case are shown in Figure~\ref{fig:E_ProblemIllustration1}. The upper plot shows the power optimization objectives, and the lower plot shows the torque gain optimization parameter. For both objectives, the dither excitation signal frequency is chosen equal at a frequency below the system's bandwidth of the input to aerodynamic power objective at ${\omega\_{d}=0.1}$~rad/s, and the demodulation phase compensation is set to ${\psi=0}$~deg. Results show that the former-mentioned objective converges to its optimal value, whereas using generator power does converge but surpasses the optimal value because the phase angle $\psi$ is not tuned properly. 
\begin{figure}[t!]
    \begin{subfigure}[b]{\linewidth}
        \centering
        \includegraphics{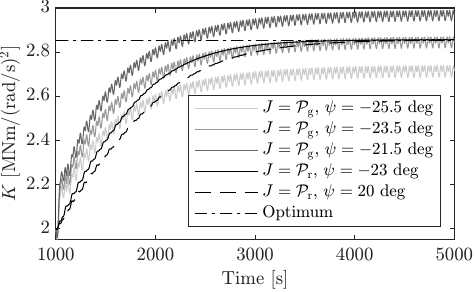}
        \caption{}
        \label{fig:E_ProblemIllustration2}
    \end{subfigure}
    \begin{subfigure}[b]{\linewidth}
        \centering
        \includegraphics{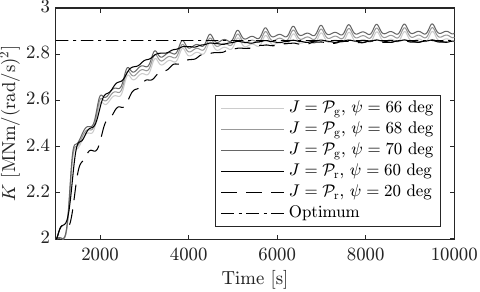}
        \caption{}
        \label{fig:E_ProblemIllustration3}
    \end{subfigure}
    \caption{Torque gain optimization on aerodynamic power ($J=\mathcal{P}_\mathrm{r}$) and generator power ($J=\mathcal{P}_\mathrm{g}$) objectives, with demodulation phase compensation. Plots (a) and (b) respectively show the results for dither frequencies $\omega\_{d}=0.1$~rad/s and $\omega\_{d}=0.01$~rad/s. The final convergence value of ESC based on the generator power objective is more sensitive to the chosen demodulation phase when the dither frequency is chosen as $0.1$~rad/sec. An arbitrary phase offset is included for aerodynamic power (\texttt{--}) to show that the phase offset only influences the convergence rate.}
    \label{fig:E_ProblemIllustration23}
\end{figure}

The results for the second case are presented in Figure~\ref{fig:E_ProblemIllustration2}, where only the torque gain trajectory is shown. The dither frequency of ${\omega\_{d}=0.1}$~rad/s retained; however, demodulation phase compensation is now used. The demodulation phase compensates for the phase loss induced by the combined system and high-pass filter dynamics ($\psi=-23$~deg) and improves convergences for the aerodynamic power case. For the generator power objective, a unique phase compensation angle converged to the actual optimum. Perturbing the phase angle slightly results in biased convergence.

To explore whether the dither frequency might be chosen too high for the generator power objective, the third case lowers the dither frequency by an order of magnitude to $0.01$~rad/s for both objectives. Results are shown in Figure~\ref{fig:E_ProblemIllustration3} and indicate that the convergence sensitivity to the specific value of the phase compensation angle is reduced.

A detailed analysis of the underlying cause of this convergence sensitivity phenomenon, and how it may be overcome, is the concern of the rest of this paper.

\section{DYNAMIC ANALYSIS}\label{sec:DA}
\noindent Throughout this section, we present a qualitative interpretation to explain the phenomena described in Section~\ref{sec:PI}. Specifically, we analyze successive linearizations of the system $\mathcal{P}$, evaluated in the frequency domain.
%This section analyses the dynamics of the system $\mathcal{P}$. A nonlinear state space model based on the definitions in Section~\ref{sec:T} is defined, and linearizations thereof are analyzed in the frequency domain.

\subsection{Derivation}
\begin{figure*}[t!]
    \centering
    \begin{subfigure}[b]{0.48\textwidth}
        \centering
        \includegraphics{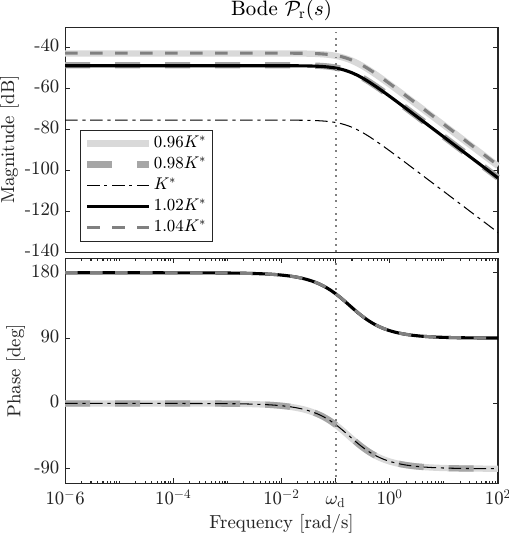}
        \caption{}
        \label{fig:P_BodePrs}
    \end{subfigure}
    \hfill
    \begin{subfigure}[b]{0.48\textwidth}
        \centering
         \includegraphics{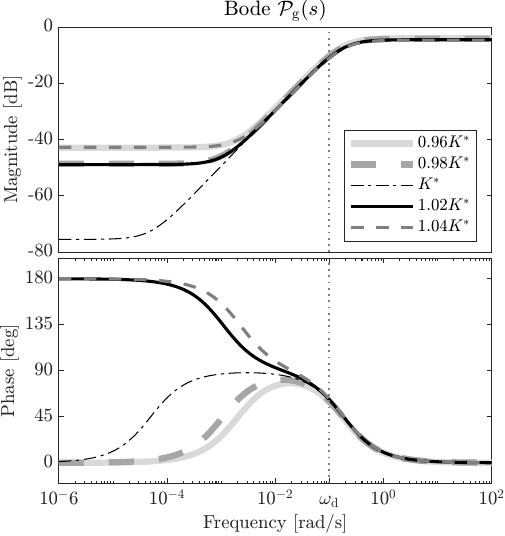}
        \caption{}
        \label{fig:P_BodePgs}
    \end{subfigure}
    \caption{Bode plots of the transfer functions $\mathcal{P}\_{r}$ from torque gain input $K$ to aerodynamic power (a), and transfer functions from the same input to generator power (b). The frequency responses represent the dynamics around the optimal torque gain value $K^*$. The dynamics of both systems differ by introducing a zero for the generator power objective. This zero crosses the imaginary axis when surpassing $K^*$, transitioning the system from minimum- to nonminimum-phase. Whereas $\mathcal{P}\_{r}$ shows a consistent $180$~deg phase difference around $K^*$, this phase difference is only present at lower frequencies of $\mathcal{P}\_{g}$. This impacts ESC based on generator power and makes the final convergence value dependent on the demodulation phase compensation angle $\psi$ when the dither frequency is chosen too high.}
    \label{fig:P_BodePrgs}
\end{figure*}
\begin{figure}[t!]
    \centering
    \includegraphics{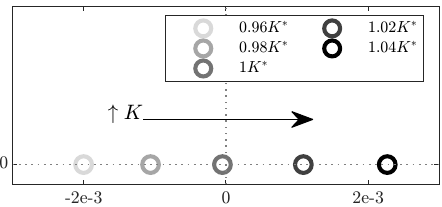}
    \caption{A zero-plot of the transfer function $\mathcal{P}\_{g}(s)$ for increasing torque gains crossing the optimal value. Using the generator power as an objective introduces a zero from the input $K$ to the output $P\_{g}$. The system alternates between a minimum- and nonminimum-phase system around $K^{*}$.}
    \label{fig:P_ZeroPlotPg}
\end{figure}
\noindent The nonlinear state-space system is defined by taking the state as~$x = \left.\omega\_{r}\right.$, the input $u = \left.K\right.$, and the outputs as $y = \left[P\_{r},\,P\_{g}\right]^\top$ resulting in the state and output equations:
\begin{align}
    \dot{x} &= f(x,u) = \frac{1}{I}\left(\tau\_{r} - \tau\_{g}\right),\label{eq:DA_fxu}\\
    y &= g(x,u) = \begin{cases}
      P\_{r} &= \frac{1}{2}\rho A C\_{p}(\lambda) V^3\\
      P\_{g} &= K\omega\_{r}^3
    \end{cases}.\label{eq:DA_gxu}
\end{align}
Linearizing the above-given system for both objectives gives the state, input, output, and direct feedthrough matrices
\begin{subequations}
    \begin{align}\label{eq:lin:AB}
        A &= \pderiv{f(x,u)}{x} = \frac{1}{I}\left(\frac{1}{2}\rho AR^2V \pderiv{C\_{\tau}(\bar{\lambda})}{\lambda} - 2K\bar{\omega}\_{r} \right),\\
        B &= \pderiv{f(x,u)}{u} = -\bar{\omega}\_{r}^2/I,\\
        C &= \pderiv{g(x,u)}{x} = \begin{bmatrix} C_1 \\ C_2 \end{bmatrix} =  \begin{bmatrix}
            \frac{1}{2}\rho A RV^2\pderiv{C\_{p}(\bar{\lambda})}{\lambda} \\
            3K\bar{\omega}\_{r}^2
        \end{bmatrix},\\
        D &= \pderiv{g(x,u)}{u} = \begin{bmatrix}
            0\\
            \bar{\omega}\_{r}^3
        \end{bmatrix}.
    \end{align}
\end{subequations}
The transfer function matrix of the state-space system is computed using ${C(sI-A)^{-1}B+D}$ for convenient analytical evaluation
\begin{align}
    \mathcal{P}(s) = \begin{bmatrix} \mathcal{P}\_{r}(s) \\  \mathcal{P}\_{g}(s) \end{bmatrix} = \begin{bmatrix} {\mathcal{N}\_{r,1}}/{\mathcal{D}(s)} \\ ({\mathcal{N}\_{g,2}s}+{\mathcal{N}\_{g,1}})/{\mathcal{D}(s)}\end{bmatrix},\label{eq:DA_TFs}
\end{align}
with 
\begin{align*}
    \mathcal{N}\_{r,1} &= -\frac{1}{2}\rho ARV^2\pderiv{C\_{p}(\bar{\lambda})}{\lambda}\bar{\omega}\_{r}^{2},\\
    \mathcal{N}\_{g,1} &= -\frac{1}{2}\rho AR^{2}V\pderiv{C\_{\tau}(\bar{\lambda})}{\lambda}\bar{\omega}\_{r}^{3}-K\bar{\omega}\_{r}^{4},\\
    \mathcal{N}\_{g,2} &= I\bar{\omega}\_{r}^{3},\\
    \mathcal{D}(s) &= Is+\left(2K\bar{\omega}\_{r}-\frac{1}{2}\rho AV\pderiv{C\_{\tau}(\bar{\lambda})}{\lambda}R^2\right).
\end{align*}
The two systems have equal system characteristics by the transfer function denominator $\mathcal{D}(s)$. However, they differ in the I/O dynamics by distinct numerator expressions. The next section further elaborates.

\subsection{Frequency-domain evaluation}
\noindent This section presents a frequency domain analysis of the derived SIMO system. To this end, the linear system is evaluated at steady-state operating conditions resulting from torque gain values around the optimum $K^*$ and a fixed wind speed of $\bar{V} = 8$~m/s. Specifically, the system is evaluated in the neighborhood of the optimal torque gain at the operating conditions belonging to the set ${K \in \left\{ 0.96,\, 0.98,\, 1.0,\, 1.02,\, 1.04 \right\}}\times K^{*}$.  

Figure~\ref{fig:P_BodePrgs} shows the frequency responses side-by-side for both I/O pairs for a frequency range ${\Omega = \left\{\omega\in\mathbb{R}\,|\,10^{-6}\leq\omega\leq10^{2}\right\}}$. The plots are read as follows: the frequency-axis represents the dither frequency at which the system is excited. The periodic dither signal is applied to $K$ and -- together with the slower time-varying ESC contribution $\tilde{K}$ -- directly affects the system dynamics. Therefore, the plots need to be read for a specific dither frequency $\omega\_{d}$ and over the frequency responses related to the torque gains. As the input $K$ varies the system dynamics, care has to be taken to interpret the results, especially at the optimal value $K^*$. However, the results form intuition and provide insights into the underlying dynamics of the system around the convergence value.

Starting from the comparison of the frequency response phase plots of $\mathcal{P}\_{r}(s)$ with $\mathcal{P}\_{g}(s)$ below ${(K<K^{*})}$ and above ${(K>K^{*})}$, some effects can be derived regarding the distinct convergence behavior. Whereas $\mathcal{P}\_{r}(s)$ shows a $180^\circ$ phase change (sign-flip) over the whole frequency range, once the optimal value $K^*$ has been crossed, the phase plot of $\mathcal{P}\_{g}(s)$ only demonstrates this behavior in the lower range of the interval $\Omega$. The sign-flipping phase behavior is required for a dither-demodulation ESC scheme to converge and can be understood in the time domain, as explained in~\cite{Brunton2022Book}. 

Apart from the analysis of the phase plot of the transfer function, it can be seen from~\eqref{eq:DA_TFs} that there is an additional zero that is introduced for $\mathcal{P}\_{g}(s)$, compared to $\mathcal{P}\_{r}(s)$. This zero makes the system transition between minimum and nonminimum-phase, around the optimal value $K^*$. Figure~\ref{fig:P_ZeroPlotPg} presents a zero map of $\mathcal{P}\_{g}(s)$ illustrating this behavior.

Thus, while $\mathcal{P}\_{r}(s)$ directly provides favorable phase behavior by a sign flip in the numerator $\mathcal{N}\_{r,1}$ when crossing $K^{*}$, for the transfer function $\mathcal{P}\_{g}(s)$, this phase behavior is affected by the introduced zero. At low frequencies, a $180^\circ$ phase difference is observed. Whenever the zero takes effect, this results in a $90^\circ$ phase increase for ${(K<K^{*})}$, and a $-90^\circ$ phase drop for ${(K>K^{*})}$, the latter of which is caused by nonminimum-phase effects. For this mid-frequency region of $\Omega$, the phase responses are all around $90^\circ$, and the system dynamics (pole) results in all phase trajectories arriving at $0^\circ$. 

For generator-based ESC, selecting the dither frequency in a higher frequency region without enough phase difference over the optimal value $K^{*}$ requires proper tuning of the demodulation phase angle $\psi$.
%and is the underlying cause of the observations in Section~\ref{sec:PI}. 
Convergence towards the actual optimal value based on generator power with reduced sensitivity to demodulation phase compensation is possible by selecting lower dither frequencies, as the phase difference increases in this direction. The following section proposes a first solution to allow the selection of higher excitation frequencies, also leading to faster convergence.

\section{PROPOSED SOLUTION}\label{sec:S}
\noindent This section proposes a solution to overcome the challenges of generator power-based ESC by reconstructing an estimate of the aerodynamic power using rotor speed acceleration, such that 
\begin{equation}\label{eq:S_Prhat}
    \hat{P}\_{r}(\dot{\hat{\omega}}\_{r}) = I \omega_r \dot{\hat{\omega}}\_{r} + P\_{g},
\end{equation}
in which $\dot{\hat{\omega}}\_{r}$ is the estimated rotor acceleration. This objective is defined as the output function $\hat{g}(x,u) = \hat{P}\_{r}$.
%included as an additional output in~\eqref{eq:DA_gxu} such that an augmented output vector is formed:
% \begin{align*}
%     g^{+}(x,u) &= \left[g(x,u)^{\top}\,\hat{P}\_{r}\right]^{\top}\label{eq:S_gxu_aug}.
% \end{align*} 

To show the potential of this solution, at first a perfect acceleration signal is assumed to be available, after which this assumption is relaxed, by introducing an estimate of the acceleration signal via filtered numerical differentiation.

\begin{remark}
    A more noise-resilient method uses a state observer to obtain an acceleration estimate. However, an inaccurate system model for the observer leads to inaccurate acceleration transient estimates and biased ESC convergence. That is, the ESC algorithm would converge but to a non-optimal value of the actual turbine. Therefore, also to retain the model-free approach, this section proposes a numerical approach to circumvent the need for an -- often unknown or inaccurate -- system model. The proposed solution exposes the solution's potential when the actual rotor acceleration is known.
\end{remark}

\subsection{Perfect acceleration measurement}\label{sec:S_Ideal}
\noindent Suppose an exact measurement of the angular acceleration $\dot{\hat{\omega}}\_{r} = \dot{\omega}\_{r}=\dot{x}$ is available. Then, evaluating~\eqref{eq:S_Prhat} results in
\begin{align}
    \begin{split}
    \hat{P}\_{r}(\dot{\omega}\_{r}) &= I\omega\_{r}\dot{\omega}\_{r} + P_g = \\
    &= I\omega\_{r}\left(\frac{1}{I}\left(\frac{1}{2}\rho A R V^2 C\_{\tau} - K\omega\_{r}^2\right)\right) + K\omega\_{r}^3\\
    &= \frac{1}{2}\rho AV^3 C\_{p}(\lambda) = P\_{r}.
    \end{split}
\end{align}
In this idealized scenario, with perfect knowledge of the rotor acceleration, the estimated aerodynamic power equals the actual aerodynamic power. The next section considers obtaining a more realistic approach to obtaining the derivative.
% Consequently, linearizing \eqref{eq:S_gxu_aug} with the same state $x = \omega_r$ and input $u = K$ as before results in:
% \begin{align}
%     C^{+} = \pderiv{g^{+}(x,u)}{x} = \begin{bmatrix}C_1 \\ C_2 \\ C_1 \end{bmatrix},~~D^{+} = \pderiv{g^{+}(x,u)}{u} = \begin{bmatrix}D_1 \\ D_2 \\ D_1 \end{bmatrix},\nonumber
% \end{align}
% confirming that the first and third outputs are equal, while matrices $A$ and $B$ are the same as per their definition in \eqref{eq:lin:AB} as the state equation remained unchanged.

\subsection{Filtered numerical differentiation}\label{sec:S_NUMDERIV}

\noindent In practice, the rotor speed acceleration could be obtained by numerical differentiation of the speed signal. This section employs a filtered numerical derivative~\cite{IEEE2016_Std421} that is given in the following differential form:
\begin{align}
    \dot{x}\_{d} &= \frac{1}{T\_{d}}(K\_{d}u\_{d}-x\_{d}),\label{eq:S_FiltDeriv}
    % &y = \dot{\hat{\omega}}\_{r} = \frac{1}{T\_{d}}(K\_{d}u\_{d}-\hat{\omega}\_{r}),
\end{align}
where $T\_{d}$ is the filter time constant and $K\_{d}$ the filter gain, and the internal state of the filter is $x\_{d} = \hat{\omega}\_{r}$ with the input being the measured rotor speed $u\_{d} = \omega\_{r}$. The time constant determines the pole location of the filter to limit the amplification of high-frequency noise. The quality of the resulting numerical derivative depends on the chosen filter time constant: choosing the value too low results in the amplification of high-frequency noise, whereas a higher value excludes information of higher frequency components.
\begin{figure}[t!]
    \centering
    \includegraphics{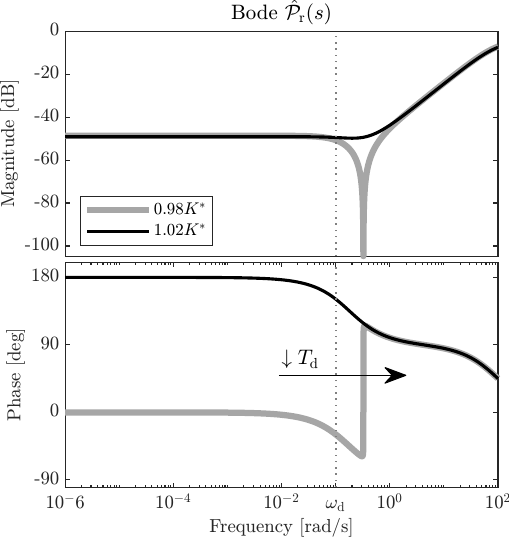}
    \caption{Bode plot of the system with an augmented filtered derivative state equation for obtaining an estimate of the rotor acceleration, and resultingly, an aerodynamic power estimate. Properly selecting the filter time constant $T\_{d}$ results in beneficial phase properties. This allows for selecting higher dither frequencies and improves ESC convergence.}
    \label{fig:P_BodePrHats}
\end{figure}

Using~\eqref{eq:S_Prhat}~and~\eqref{eq:S_FiltDeriv}, the following expression for the estimated aerodynamic power is obtained
\begin{align}
    \hat{P}\_{r}(\dot{\hat{\omega}}\_{r}) = I\omega_r \left(\frac{1}{T\_{d}}(K_d \omega_r - \hat{\omega}_r)\right) + P_g,
\end{align}
which can then be linearized, now with the augmented state ${\hat{x} = \left[\omega\_r,\,\hat{\omega}\_r\right]^\top}$. Because of the change in the definition of the state variable, the linear state space matrices are redefined as: 
\begin{align*}
    \begin{split}
        &\hat{A} = \begin{bmatrix}
            A &0 \\ {K\_{d}}/{T\_{d}} &-{1}/{T\_{d}}
        \end{bmatrix},\quad \hat{B} = \begin{bmatrix}
            B \\ 0
        \end{bmatrix} \\
        &\hat{C} = \begin{bmatrix}
            \pderivil{\hat{P}\_{r}}{\omega\_{r}} & \pderivil{\hat{P}\_{r}}{\hat{\omega}\_{r}}
            \end{bmatrix},\quad
        \hat{D} = \begin{bmatrix}
            \pderivil{\hat{P}\_{r}}{K}
        \end{bmatrix},
    \end{split}
\end{align*}
where the partial derivatives in $\hat{C}$ and $\hat{D}$ are:
\begin{align*}
    &\pderiv{\hat{P}\_{r}}{\omega\_{r}} = \frac{I}{T\_{d}}\left(2 K\_{d} \bar\omega - \bar{\hat{\omega}}\_{r}\right) + 3 \bar{K}\bar{\omega}\_{r}^2,\\
    &\pderiv{\hat{P}\_{r}}{\hat{\omega}\_{r}} = - \frac{I}{T\_{d}}\bar{\omega}\_{r},\qquad \pderiv{\hat{P}\_{r}}{K} = \bar{\omega}\_{r}^3.
\end{align*}
Figure~\ref{fig:P_BodePrHats} presents a Bode plot of the estimated system at operating conditions just below and above the optimal value. Magnitude and phase frequency responses are shown for a gain value ${K\_{d}=1}$ and a filter time constant of ${T\_{d}=10^{-2}}$~s. Comparing with the Bode plots in Figures~\ref{fig:P_BodePrs}~and~\ref{fig:P_BodePgs} shows that for the lower frequency range, the responses follow the trajectory of the formerly mentioned figure, whereas for higher frequencies (after the anti-resonance), the properties of the latter mentioned figure resemble. This behavior is an effect of the anti-resonance (complex-conjugate zero pair) for systems ${K<K^{*}}$, and results in desirable phase behavior.  Formerly, at the dither frequency of $\omega\_{d}=0.1$~rad/s in Figure~\ref{fig:P_BodePgs}, no sign-alternating phase was seen; now, with the addition of the acceleration term, this behavior is attained. 

In the limit for $T\_{d}\rightarrow0$, the filtered derivative reduces to a pure differentiator, leading to the ideal situation described in Section~\ref{sec:S_Ideal}. The filter time constant is a trade-off between noise attenuation for higher values and beneficial phase properties for lower time constant values, allowing the selection of higher dither frequencies.

\subsection{Results}\label{sec:S_Results}
\noindent This section validates the results from the analysis presented in this paper and shows the effectiveness of the proposed solution for effective generator power-based optimization by simulations. The ESC scheme takes either the actual aerodynamic power $P\_{r}$ or its estimate $\hat{P}\_{r}$ as the input objective. For the latter mentioned, the derivative time constant is taken as $T\_{d} = 10^{-2}$~s, and the convergence properties are evaluated for phase offsets $\psi=\{-30,\,0,\,30\}$~deg. The internal ESC integrator state is initialized at $\tilde{K}_0=0.7K^{*}$ with an integral gain of $\kappa =4\cdot10^{4}$. A dither amplitude of $A\_{d} = 1\cdot 10^5$~Nm/(rad/s)\ts{2} is used and the dither and demodulation signals have a frequency of $\omega\_{d}=0.1$~rad/s. The high- and low-pass filter bandwidths are taken according to~\eqref{eq:ESC_Alg_FilterW}.

\begin{figure}[t!]
    \centering
    \includegraphics{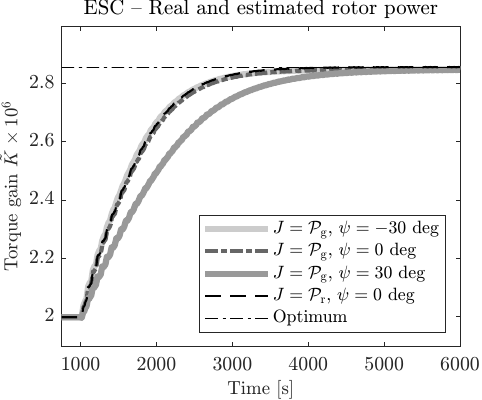}
    \caption{Torque gain convergence to the optimal value $\tilde{K}^*$ using ESC convergence at equal dither frequencies for both the measured actual and estimated aerodynamic power. For clarity of presentation, the internal ESC integrator state $\tilde{K}$ is displayed without adding the scaled dither signal $d_1(t)$.}
    \label{fig:S_ResultsSimulationPrHat}
\end{figure}

Figure~\ref{fig:S_ResultsSimulationPrHat} shows the ESC optimization torque gain results. After initialization, the ESC algorithm is enabled after $1000$~s. For clarity, only the evolution of the ESC internal state $\tilde{K}$ is presented without the added periodic dither contribution. A direct comparison can be made to Figure~\ref{fig:E_ProblemIllustration2}: Using the estimated aerodynamic power for the various phase offsets, the convergence characteristics towards the actual optimal value are now consistent and only impact the convergence speed.

\section{CONCLUSIONS}\label{sec:C}
\noindent An analysis has been performed on using aerodynamic and generator power measurements as objectives in dither-demodulation extremum seeking control for wind turbine torque controller optimization. It was found that the dynamics differ for the system under optimization with torque gain input and distinct objectives as outputs. The input to the generator power objective system turns from minimum- to nonminimum-phase when crossing the optimal torque controller calibration. This impacts the desirable phase behavior needed for ESC convergence at higher dither excitation frequencies, which may not be appropriate for generator-based ESC convergence. Lowering the dither frequency would enable convergence but reduce the convergence rate. A solution has been proposed by forming a new estimated aerodynamic power objective augmenting the measured generator power with rotor acceleration dynamics. By obtaining the acceleration through filtered numerical derivative, results show faster dither is allowed, also leading to improved convergence rates.

\bibliographystyle{IEEEtran} 
\bibliography{references}

\begin{thebibliography}{10}
\providecommand{\url}[1]{#1}
\csname url@rmstyle\endcsname
\providecommand{\newblock}{\relax}
\providecommand{\bibinfo}[2]{#2}
\providecommand\BIBentrySTDinterwordspacing{\spaceskip=0pt\relax}
\providecommand\BIBentryALTinterwordstretchfactor{4}
\providecommand\BIBentryALTinterwordspacing{\spaceskip=\fontdimen2\font plus
\BIBentryALTinterwordstretchfactor\fontdimen3\font minus
  \fontdimen4\font\relax}
\providecommand\BIBforeignlanguage[2]{{%
\expandafter\ifx\csname l@#1\endcsname\relax
\typeout{** WARNING: IEEEtran.bst: No hyphenation pattern has been}%
\typeout{** loaded for the language `#1'. Using the pattern for}%
\typeout{** the default language instead.}%
\else
\language=\csname l@#1\endcsname
\fi
#2}}
\renewcommand\BIBentryALTinterwordstretchfactor{4}

\bibitem{Brandetti2022}
L.~Brandetti, Y.~Liu, S.~P. Mulders, C.~Ferreira, S.~Watson, and J.~W. {van
  Wingerden}, ``{On the ill-conditioning of the combined wind speed estimator
  and tip-speed ratio tracking control scheme},'' \emph{Journal of Physics:
  Conference Series}, 2022.

\bibitem{Johnson2006}
K.~E. Johnson, L.~Y. Pao, M.~J. Balas, and J.~F. Lee, ``{Control variable-speed
  wind turbines: Standard and adaptive techniques for maximizing energy
  capture},'' \emph{IEEE Control Systems}, 2006.

\bibitem{Mulders2023_ACC}
S.~P. Mulders, L.~Brandetti, F.~Spagnolo, Y.~Liu, P.~B. Christensen, and J.~W.
  van Wingerden, ``A learning algorithm for the calibration of internal model
  uncertainties in advanced wind turbine controllers: {A} wind speed
  measurement-free approach,'' in \emph{American Control Conference (ACC)},
  2023.

\bibitem{Mulders2023_IFAC}
S.~P. Mulders, Y.~Liu, F.~Spagnolo, P.~Christensen, and J.~W. van Wingerden,
  ``An iterative data-driven learning algorithm for calibration of the internal
  model in advanced wind turbine controllers,'' in \emph{IFAC World Congress},
  2023.

\bibitem{Brandetti2023_MOControlCalibration}
L.~Brandetti, S.~P. Mulders, Y.~Liu, S.~Watson, and J.~W. van Wingerden,
  ``Analysis and multi-objective optimisation of model-based wind turbine
  controllers,'' \emph{Wind Energy Science}, 2023.

\bibitem{Ariyur2003_ESCBook}
K.~B. Ariyur and M.~Krstic, \emph{Real-time optimization by extremum-seeking
  control}.\hskip 1em plus 0.5em minus 0.4em\relax John Wiley \& Sons, 2003.

\bibitem{Rotea2001_ACC}
M.~Rotea, ``Analysis of multivariable extremum seeking algorithms,'' in
  \emph{American Control Conference (ACC)}, 2000.

\bibitem{Tan2010_ESC1922to2010}
Y.~Tan, W.~Moase, C.~Manzie, D.~Nešić, and I.~Mareels, ``Extremum seeking
  from 1922 to 2010,'' in \emph{Proceedings of the 29th Chinese Control
  Conference}, 2010.

\bibitem{Komatsu2001_ESCFinePitch}
M.~Komatsu, H.~Miyamoto, H.~Ohmori, and A.~Sano, ``Output maximization control
  of wind turbine based on extremum control strategy,'' in \emph{American
  Control Conference}, 2001, pp. 1739--1740 vol.2.

\bibitem{Ishii2003_ESCMicroWindTubine}
C.~Ishii, H.~Hashimoto, and H.~Ohmori, ``Modeling of variable pitch micro wind
  turbine and its output optimization control with adaptive extremum control
  scheme,'' \emph{Nippon Kikai Gakkai Ronbunshu}, 2003.

\bibitem{Creaby2009_MultiVarESC}
J.~Creaby, Y.~Li, and J.~E. Seem, ``Maximizing wind turbine energy capture
  using multivariable extremum seeking control,'' \emph{Wind Engineering},
  2009.

\bibitem{Xiao2016_CART3Sim}
Y.~Xiao, Y.~Li, and M.~Rotea, ``Experimental evaluation of extremum seeking
  based region-2 controller for {CART3} wind turbine,'' in \emph{34th Wind
  Energy Symposium}, 2016, p. 1737.

\bibitem{Xiao2018_CART3Field}
Y.~Xiao, Y.~Li, and M.~A. Rotea, ``{CART3} field tests for wind turbine
  region-2 operation with extremum seeking controllers,'' \emph{IEEE
  Transactions on Control Systems Technology}, 2018.

\bibitem{Rotea2017_LogofPowers}
M.~A. Rotea, ``Logarithmic power feedback for extremum seeking control of wind
  turbines.''\hskip 1em plus 0.5em minus 0.4em\relax Elsevier, 2017.

\bibitem{Ciri2019_LESLPESC}
U.~Ciri, S.~Leonardi, and M.~A. Rotea, ``Evaluation of log-of-power extremum
  seeking control for wind turbines using large eddy simulations,'' \emph{Wind
  Energy}, 2019.

\bibitem{Kumar2022LPPIESC}
D.~Kumar and M.~A. Rotea, ``Wind turbine power maximization using log-power
  proportional-integral extremum seeking,'' \emph{Energies}, vol.~15, no.~3,
  2022.

\bibitem{Brunton2022Book}
S.~L. Brunton and J.~N. Kutz, \emph{Data-driven science and engineering:
  Machine learning, dynamical systems, and control}.\hskip 1em plus 0.5em minus
  0.4em\relax Cambridge University Press, 2022.

\bibitem{IEEE2016_Std421}
{IEEE}, ``{IEEE} recommended practice for excitation system models for power
  system stability studies,'' \emph{{IEEE} {S}td 421.5-2016}, 2016.

\end{thebibliography}

\end{document}